\documentclass[aps,prd,twocolumn,eqsecnum,showpacs,amsmath]{revtex4}

\usepackage{subfigure}
\usepackage[dvips]{color,graphicx}
\usepackage{amsfonts,amssymb,theorem}
{\theorembodyfont{\upshape}
}
{\theorembodyfont{\upshape}
}
{\theorembodyfont{\upshape}
}
{\theorembodyfont{\upshape}
}
{\theorembodyfont{\upshape}
}
{\theorembodyfont{\upshape}
}

\newcommand{\dalm}{\kern1pt\vbox{\hrule height 0.9pt\hbox{\vrule width
0.9pt\hskip 2.5pt\vbox{\vskip 5.5pt}\hskip 3pt\vrule width 0.3pt}\hrule height
0.3pt}\kern1pt}

\def\b2hat{ {\hat b}_2 }

\begin{document}

\title{
Hedgehog ansatz and its generalization for self-gravitating Skyrmions
}

\author{Fabrizio Canfora${}^{a,b}$}
\email{canfora-at-cecs.cl}
\author{Hideki Maeda${}^{a}$}
\email{hideki-at-cecs.cl}


\address{ 
	${}^a$ Centro de Estudios Cient\'{\i}ficos (CECs), Casilla 1469,
 Valdivia, Chile \\
	${}^b$ Universidad Andres Bello, Av. Republica 440, Santiago, Chile
}

\date{\today}

\begin{abstract} 
The hedgehog ansatz for spherically symmetric spacetimes in self-gravitating nonlinear sigma models and Skyrme models is revisited and its generalization for non-spherically symmetric spacetimes is proposed.
The key idea behind our construction is that, even if the matter fields depend on the Killing coordinates in a nontrivial way, the corresponding energy-momentum tensor can still be compatible with spacetime symmetries.
Our generalized hedgehog ansatz reduces the Skyrme equations to coupled differential equations for two scalar fields together with several constraint equations between them.
Some particular field configurations satisfying those constraints are presented in several physically important spacetimes, including stationary and axisymmetric spacetimes.
Incidentally, several new exact solutions are obtained under the standard hedgehog ansatz, one of which represents a global monopole inside a black hole with the Skyrme effect.
\end{abstract}

\maketitle


\section{Introduction}

Nonlinear sigma models are among the most important nonlinear field theories
due to their many applications, ranging from quantum field theory to
statistical mechanics. (See Ref.~\cite{manton} for a detailed review.) Examples
are quantum magnetism, the quantum hall effect, mesons, and string theory. It
has also been successfully applied as an effective field theory to super
fluid ${}^{3}$He. A sigma model in $D$-dimensional spacetime $(\mathcal{%
M}^{D},g_{\mu \nu })$ is defined by a set of $n$ real scalar fields $%
Y^{i}~(i=1,\cdots ,n)$ which take on values in a flat manifold, called the
target manifold. It is called a nonlinear sigma model if the target manifold
is non-flat, Lagrangian density of which is given by 
\begin{equation*}
\mathcal{L}=\frac{1}{2}g^{\mu \nu }G_{ij}(\nabla _{\mu }Y^{i})(\nabla _{\nu
}Y^{j}),
\end{equation*}%
where $G_{ij}(Y)$ is the metric on the target manifold.

Actually, nonlinear sigma models do not admit any static soliton solutions
in 3+1 dimensions, which is shown by a scaling argument. (See Ref.~\cite{manton}
for instance). For this reason, Skyrme introduced his famous term, which
allows the existence of static solutions with finite energy called \textit{%
Skyrmions}~\cite{skyrme}. Remarkably, excitations around Skyrme solitons may
represent Fermionic degrees of freedom suitable to describe nucleons. The
Skyrme model is therefore one of the most important nonlinear field theories
in nuclear and high-energy physics.

However, it is difficult to obtain exact solutions in nonlinear sigma models
or Skyrme models, due to their highly nonlinear characters. Therefore one
often adopts a certain ansatz to make the field equations more tractable.
Under such ans{\"a}tze, the results can be interpreted more clearly and the
simplified equations are also useful for numerical studies. Among others,
the best known one for Skyrme models is the hedgehog ansatz for spherically
symmetric systems, which reduces the field equations to a single scalar
equation.

Because of its great advantage, the hedgehog ansatz has been also adopted in
self-gravitating Skyrme models. The Einstein-Skyrme system has attracted
considerable attention since Droz, Heusler, and Straumann numerically found
spherically symmetric black-hole solutions with a nontrivial Skyrme field,
namely a Skyrme hair~\cite{dhs1991a}. (Before them, Luckock and Moss
numerically constructed such hairy configurations in the Schwarzschild
background spacetime~\cite{lm1986}.) This was the first counterexample to
the black hole no-hair conjecture, and it is stable against spherical linear
perturbations~\cite{dhs1991b}. Regular particle-like configurations~\cite%
{SkyrmeRegular} and dynamical properties of the system have also been
investigated numerically~\cite{Skyrmedynamics}.

In this decade, not only spherically symmetric configurations~\cite%
{SkyrmeSpherical} but also more realistic black holes or regular
configurations with axisymmetry have been studied in the Einstein-Skyrme
system~\cite{SkyrmeAxi}. In those studies, one mostly relies on numerical
analyses because of the complexity of the system. (See Ref.~\cite{skyrmeBHreview}
for a review.) Under these circumstances, it would be helpful for both
analytic and numerical investigations to provide a new useful ansatz which
also makes the field equations much simpler and tractable. In the present
paper, we generalize the hedgehog ansatz in an applicable way not only to
spherically symmetric spacetimes but also to other symmetric spacetimes.

In the following section, we review the Einstein-Skyrme system in the
presence of a cosmological constant. In Sec.~\ref{sec3}, we revisit the
standard hedgehog ansatz in spherically symmetric spacetimes and obtain a
new exact black-hole solution. In Sec.~\ref{sec4}, we present the
generalized hedgehog ansatz and derive the basic equations. We also present
some particular configurations which are compatible with a variety of
symmetric spacetimes. Concluding remarks and future prospects are summarized
in Sec.~\ref{sec5}. Our basic notation follows Ref.~\cite{wald}. The conventions
for curvature tensors are $[\nabla _{\rho },\nabla _{\sigma }]V^{\mu }={%
\mathcal{R}^{\mu }}_{\nu \rho \sigma }V^{\nu }$ and $\mathcal{R}_{\mu \nu }={%
\mathcal{R}^{\rho }}_{\mu \rho \nu }$. The signature of the Minkowski
spacetime is $(-,+,+,+)$ and Greek indices run over all spacetime indices.
We adopt the units such that $c=\hbar =1$.


\section{The Einstein-Skyrme system}

In the present paper, we study the Einstein-Skyrme system with a
cosmological constant $\Lambda $ in four dimensions. A Skyrme field is
described by a nonlinear sigma model with additional terms and can be
conveniently written in terms of an SU(2) group-valued scalar field $U$. The
dynamical sector in the total action of this system is written as~\cite%
{skyrmeBHreview} 
\begin{equation}
S=S_{\mathrm{G}}+S_{\mathrm{Skyrme}},
\end{equation}%
where the gravitational action $S_{\mathrm{G}}$ and the Skyrme action $S_{%
\mathrm{Skyrme}}$ are given by 
\begin{align}
S_{\mathrm{G}}=& \frac{1}{16\pi G}\int d^{4}x\sqrt{-g}(\mathcal{R}-2\Lambda
), \\
S_{\mathrm{Skyrme}}=& \int d^{4}x\sqrt{-g}\mathrm{Tr}\left( \frac{F_{\pi
}^{2}}{16}R^{\mu }R_{\mu }+\frac{1}{32e^{2}}F_{\mu \nu }F^{\mu \nu }\right)
\ .  \label{sky}
\end{align}%
Here $R_{\mu }$ and $F_{\mu \nu }$ are defined by 
\begin{align}
R_{\mu }:=& U^{-1}\nabla _{\mu }U\ ,  \label{skyrme2} \\
F_{\mu \nu }:=& \left[ R_{\mu },R_{\nu }\right], \   \label{skyrmenotation}
\end{align}%
while $G$ is the Newton constant and the parameters $F_{\pi }$ and $e$ are
fixed by comparison with experimental data. The first and the second terms
in $S_{\mathrm{Skyrme}}$ respectively represent a nonlinear sigma model and
the Skyrme term. Skyrme fields satisfy the dominant energy condition and the
strong energy condition~\cite{gibbons2003}.

The Skyrme Lagrangian describes the low-energy nonlinear interactions of
pions or baryons. The deep observation of Skyrme \cite{skyrme} was that if
one adds a suitable term (the Skyrme quartic term) to the Lagrangian of a
nonlinear sigma model the resulting action describes not only the low-energy
interactions of pions but also of baryons. This observation is remarkable in
that it was the first example of a purely bosonic Lagrangian able to
describe both bosons and fermions.

For convenience, defining $K:=F_{\pi }^{2}/4$ and $\lambda :=4/(e^{2}F_{\pi
}^{2})$, we write the Skyrme action as 
\begin{equation}
S_{\mathrm{Skyrme}}=\frac{K}{2}\int d^{4}x\sqrt{-g}\mathrm{Tr}\left( \frac{1%
}{2}R^{\mu }R_{\mu }+\frac{\lambda }{16}F_{\mu \nu }F^{\mu \nu }\right) .
\label{skyrme-action}
\end{equation}%
The resulting Einstein equations are 
\begin{equation}
G_{\mu \nu }+\Lambda g_{\mu \nu }=8\pi GT_{\mu \nu },
\end{equation}%
where $G_{\mu \nu }$ is the Einstein tensor and 
\begin{align}
T_{\mu \nu }=&-\frac{K}{2}\mathrm{Tr}\biggl[\biggl(R_{\mu }R_{\nu }-\frac{1}{2}g_{\mu \nu }R^{\alpha }R_{\alpha }\biggl) \nonumber \\
&+\frac{\lambda }{4}\biggl(%
g^{\alpha \beta }F_{\mu \alpha }F_{\nu \beta }-\frac{1}{4}g_{\mu \nu
}F_{\alpha \beta }F^{\alpha \beta }\biggl)\biggl].  \label{timunu1}
\end{align}%
The Skyrme equations are written as 
\begin{equation}
\nabla ^{\mu }R_{\mu }+\frac{\lambda }{4}\nabla ^{\mu }[R^{\nu },F_{\mu
\nu}]=0.  \label{nonlinearsigma1}
\end{equation}

Here $R_{\mu }$ is expressed as 
\begin{equation}
R_{\mu }=R_{\mu }^{i}t_{i}
\end{equation}%
in the basis of the SU(2) generators $t^{i}$ (where the Latin index $i=1,2,3$
corresponds to the group index, which is raised and lowered with the flat
metric $\delta _{ij}$), which satisfy 
\begin{equation}
t^{i}t^{j}=-\delta ^{ij}\mathbf{1}-\varepsilon ^{ijk}t^{k}\ ,
\end{equation}%
where $\mathbf{1}$ is the identity $2\times 2$ matrix and $\varepsilon_{ijk}$
and $\varepsilon ^{ijk}$ are the totally antisymmetric Levi-Civita symbols
with $\varepsilon _{123}=\varepsilon ^{123}=1$. $t^{i}$ are related to the
Pauli matrices as $t_i=-i\sigma_i$. Using the identity 
\begin{equation}
\varepsilon ^{ijk}\varepsilon ^{mnk}=\delta _{i}^{m}\delta_{j}^{n}-\delta
_{i}^{n}\delta _{j}^{m} \ ,
\end{equation}%
we obtain the commutation relation of $R_{\mu }$, 
\begin{align}
\left[ R_{\mu },R_{\nu }\right] ^{i}=& -2 \varepsilon _{ijk} R_{\mu
}^{j}R_{\nu }^{k}\, .  \label{normalcon1}
\end{align}

Hereafter we will use the following standard parametrization of the
SU(2)-valued scalar $U(x^{\mu })$:%
\begin{equation}
U(x^{\mu })=Y^{0}\mathbf{1}+Y^{i}t_{i}\ ,\quad U^{-1}(x^{\mu })=Y^{0}%
\mathbf{1}-Y^{i}t_{i}\ ,  \label{standard1}
\end{equation}%
where $Y^{0}=Y^{0}(x^{\mu })$ and $Y^{i}=Y^{i}(x^{\mu })$ satisfy 
\begin{equation}
\left( Y^{0}\right) ^{2}+Y^{i}Y_{i}=1\ .  \label{standard3}
\end{equation}%
From the definition~(\ref{skyrme2}), $R_{\mu }^{k}$ is written as 
\begin{equation}
R_{\mu }^{k}=\varepsilon ^{ijk}Y_{i}\nabla _{\mu }Y_{j}+Y^{0}\nabla _{\mu
}Y^{k}-Y^{k}\nabla _{\mu }Y^{0}\ .  \label{standard4}
\end{equation}%
Using the quadratic combination
\begin{equation}
\mathcal{S}_{\mu \nu }:=\delta _{ij}R_{\mu }^{i}R_{\nu }^{j}=G_{ij}(Y)\nabla
_{\mu }Y^{i}\nabla _{\nu }Y^{j}\ , \label{cuadra1}
\end{equation}%
where 
\begin{equation}
G_{ij}:=\delta _{ij}+\frac{Y_{i}Y_{j}}{1-Y^{k}Y_{k}}\ ,
\end{equation}%
we obtain 
\begin{align}
\mathrm{Tr}(R_{\mu }R_{\nu })=& -2\mathcal{S}_{\mu \nu }, \\
\mathrm{Tr}(F_{\mu \alpha }F_{\nu }^{~\alpha })=& 8\mathcal{S}_{\mu \alpha }%
\mathcal{S}_{\nu }^{~\alpha }-8\mathcal{S}_{\mu \nu }\mathcal{S}.
\end{align}%
Using these results, we can write the Skyrme action (\ref{skyrme-action})
only with $Y^{i}$ as 
\begin{align}
S_{\mathrm{Skyrme}}=& -K\int d^{4}x\sqrt{-g}\biggl[\frac{1}{2}G_{ij}(\nabla
_{\mu }Y^{i})(\nabla ^{\mu }Y^{j})  \notag \\
& +\frac{\lambda }{4}\biggl\{\left( G_{ij}(\nabla _{\mu }Y^{i})(\nabla ^{\mu
}Y^{j})\right) ^{2} \nonumber \\
&-G_{ij}(\nabla _{\mu }Y^{i})(\nabla _{\nu
}Y^{j})G_{kl}(\nabla ^{\mu }Y^{k})(\nabla ^{\nu }Y^{l})\biggl\}\biggl],
\end{align}%
while the energy-momentum tensor (\ref{timunu1}) is expressed as 
\begin{align}
T_{\mu \nu }=&K\biggl[\mathcal{S}_{\mu \nu }-\frac{1}{2}g_{\mu \nu }\mathcal{S}+\lambda \biggl\{\mathcal{S}\mathcal{S}_{\mu \nu }-\mathcal{S}_{\mu \alpha }\mathcal{S}_{\nu }^{~\alpha } \nonumber \\
&-\frac{1}{4}g_{\mu \nu }(\mathcal{S}^{2}-%
\mathcal{S}_{\alpha \beta }\mathcal{S}^{\alpha \beta })\biggl\}\biggl]\ .\ 
\label{tmunu2}
\end{align}%
It is seen that the contribution of the Skyrme term to the energy-momentum
tensor is traceless (in four dimensions) and shares some characteristics of
a Yang-Mills field.

Here $G_{ij}$ is the metric corresponding to the group (target) manifold,
which is $S^{3}$ in the present case. It is worth noting here that if one
considers a configuration with vanishing $Y^{0}$, then $G_{ij}$ becomes $%
\delta _{ij}$.


\section{Hedgehog ansatz for spherically symmetric spacetimes}

\label{sec3}

\subsection{Tensorial formulation of the basic equations}

In this section we will derive the field equations under the standard
hedgehog ansatz for spherically symmetric spacetimes. The most general
metric with spherical symmetry may be written as 
\begin{equation}
ds^{2}=g_{AB}(y)dy^{A}dy^{B}+r(y)^{2}\gamma _{ab}(z)dz^{a}dz^{b},
\end{equation}%
where $g_{AB}~(A,B=0,1)$ and $y^{A}$ are the metric and coordinates on a
two-dimensional Lorentzian manifold $M^{2}$, respectively, while $\gamma
_{ab}~(a,b=2,3)$ and $z^{a}$ are the metric and coordinates on a
two-dimensional unit sphere $S^{2}$, respectively. We are going to derive
the basic equations under the hedgehog ansatz in a covariant form on $%
(M^2,g_{AB})$.

In terms of the group element $U$, the usual hedgehog ansatz reads%
\begin{equation}
U=\mathbf{1}\cos \alpha +\widehat{n}^{i}t_{i}\sin \alpha \ ,\quad U^{-1}=%
\mathbf{1}\cos \alpha -\widehat{n}^{i}t_{i}\sin \alpha \ ,
\end{equation}%
where $\widehat{n}^{i}=\widehat{n}^{i}(z)~(i=1,2,3)$ are given by 
\begin{equation}
\widehat{n}^{1}=\sin \theta \cos \phi \ ,\ \ \ \widehat{n}^{2}=\sin \theta
\sin \phi \ ,\ \ \ \widehat{n}^{3}=\cos \theta \ \label{hedgehog1.2-1}
\end{equation}%
and $\alpha =\alpha (y)$. Here we have adopted the coordinates on $%
(S^{2},\gamma_{ab})$ such that 
\begin{equation}
\gamma _{ab}dz^{a}dz^{b}=d\theta ^{2}+\sin ^{2}\theta d\phi ^{2}.
\end{equation}%
In terms of the variables $Y^{0}$ and $Y^{i}$, this ansatz corresponds to 
\begin{equation}
Y^{0}=\cos \alpha \ ,\ \ Y^{i}=\widehat{n}^{i}\sin \alpha \ .
\label{hedgehog1.2}
\end{equation}%
$\widehat{n}^{i}$ are normalized as $\delta _{ij}\widehat{n}^{i}\widehat{n}%
^{j}=1$ so as to satisfy Eq.~(\ref{standard3}). It is also possible to
define the normalized internal vectors $\widehat{n}^{i}$ by 
\begin{equation}
{\bar{D}}^{2}\widehat{n}^{i}=-2\widehat{n}^{i}\ ,  \label{hedgehog-es}
\end{equation}%
where ${\bar{D}}_{a}$ is the covariant derivative on $S^{2}$ and ${\bar{D}}%
^{2}:={\bar{D}}_{a}{\bar{D}}^{a}$. Namely, $\widehat{n}^{i}$ are the
eigenvectors of the Laplacian operator on $S^{2}$ with the eigenvalue $-2$.
They satisfy $\delta_{ij}({\bar{D}}_{a}\widehat{n}^{i})({\bar{D}}_{b}%
\widehat{n}^{j})=\gamma_{ab}$, which will be used in the following
calculations.

Let us derive the expression of the energy-momentum tensor (\ref{tmunu2}) in
a tensorial way on $M^{2}$. Using Eqs.~(\ref{standard4}) and (\ref%
{hedgehog1.2}), we obtain the following expression of $R_{\mu }^{k}$: 
\begin{align}
R_{\mu }^{k}dx^{\mu }=&(\widehat{n}^{k}D_{A}\alpha )dy^{A} \nonumber \\
&+\biggl(\sin^{2}\alpha \delta ^{sk}\varepsilon _{ijs}\widehat{n}^{i}{\bar{D}}_{a}%
\widehat{n}^{j}+\frac{1}{2}\sin \left( 2\alpha \right) {\bar{D}}_{a}\widehat{%
n}^{k}\biggl)dz^{a}\ ,  \label{R-decomp}
\end{align}%
where ${D}_{A}$ is the covariant derivative on $M^{2}$. Using Eqs.~(\ref{cuadra1}), (\ref{hedgehog1.2}), and (\ref{R-decomp}), we obtain 
\begin{align}
\mathcal{S}_{\mu \nu }dx^{\mu }dx^{\nu }=& (D_{A}\alpha
)(D_{B}\alpha)dy^{A}dy^{B}+\sin ^{2}\alpha \gamma _{ab}dz^{a}dz^{b}\ 
\label{cuadra3-2}
\end{align}%
and finally derive the energy-momentum tensor (\ref{tmunu2}) as 
\begin{widetext}
\begin{align}
T_{\mu \nu }dx^{\mu }dx^{\nu }=& K\biggl[\biggl(1+2\lambda r^{-2}\sin
^{2}\alpha \biggl)\biggl((D_{A}\alpha)(D_{B}\alpha )-\frac{1}{2}%
g_{AB}(D\alpha )^{2}\biggl)-g_{AB}r^{-2}\sin ^{2}\alpha \biggl(1+\frac{\lambda }{2}r^{-2}\sin
^{2}\alpha \biggl)\biggl]dy^{A}dy^{B}  \notag \\
& -\frac{1}{2}K\biggl((D\alpha )^{2}-\lambda r^{-4}\sin ^{4}\alpha \biggl)%
r^{2}\gamma _{ab}dz^{a}dz^{b},  \label{Tmunu}
\end{align}%
where $(D\alpha )^{2}:=g^{AB}(D_{A}\alpha )(D_{B}\alpha )$. The Einstein
equations are written down with the following expression of the Einstein
tensor with the $\Lambda $-term:
\begin{align}
(G_{\mu \nu }+\Lambda g_{\mu \nu })dx^{\mu }dx^{\nu }=& \biggl[-2\frac{%
D_{A}D_{B}r}{r}+g_{AB}\biggl(2\frac{D^{2}r}{r}-\frac{1-(Dr)^{2}}{r^{2}}%
+\Lambda \biggl)\biggl]dy^{A}dy^{B}  \notag \\
& +\frac{1}{2}\biggl(2\frac{D^{2}r}{r}-{}^{(2)}\mathcal{R}+2\Lambda\biggl) %
r^{2}\gamma _{ab}dz^{a}dz^{b},  \label{leftEFE}
\end{align}%
\end{widetext}
where ${}^{(2)}\mathcal{R}$ is the Ricci scalar on $M^{2}$ and $%
D^{2}:=D_{A}D^{A}$.

Next we derive the expression of the field equations (\ref{nonlinearsigma1}%
). Using the formula $\nabla^\mu u_\mu=D^Au_A+r^{-2}{\bar D}^a
u_a+2r^{-1}(D^A r)u_A$, we obtain the divergence of $R_{\mu }^{k}$ as 
\begin{align}
\nabla^\mu R_{\mu }^{k}=&\widehat{n}^{k}\biggl[D^2\alpha -r^{-2}\sin \left(
2\alpha \right)+2r^{-1}(D^Ar)(D_{A }\alpha)\biggl],  \label{hedgehog1.5-d2}
\end{align}%
where we have used Eq.~(\ref{hedgehog-es}). Hence, the field equations (\ref%
{nonlinearsigma1}) without the Skyrme term reduce to the following single
scalar equation on $M^2$: 
\begin{equation}
D^2\alpha +2r^{-1}(D^Ar)(D_{A }\alpha)-r^{-2}\sin \left( 2\alpha \right)=0\ .
\label{sundaba2.2}
\end{equation}%
This is a very nontrivial characteristic of the hedgehog ansatz, which
reduces a system of coupled nonlinear partial differential equations (\ref%
{nonlinearsigma1}) to a single equation (\ref{sundaba2.2}). Actually, this
still holds even with the Skyrme term, as shown below.

It is straightforward to show that
\begin{equation}
\left[ R^{\nu },F_{\mu \nu }\right] =4\left( \mathcal{S}R_{\mu }^{k}-%
\mathcal{S}_{~\mu }^{\nu }R_{\nu }^{k}\right) t_{k}\ .  \label{preskyrme3}
\end{equation}%
Using the two expressions
\begin{widetext}
\begin{align}
\mathcal{S}R_{\mu }^{k}dx^{\mu }=& \biggl((D\alpha )^{2}+2r^{-2}\sin
^{2}\alpha \biggl)(D_{A}\alpha )\widehat{n}^{k}dy^{A}  \notag \\
& +\biggl((D\alpha )^{2}+2r^{-2}\sin ^{2}\alpha \biggl)\biggl(\sin
^{2}\alpha \delta ^{sk}\varepsilon _{ijs}\widehat{n}^{i}{\bar{D}}_{a}%
\widehat{n}^{j}+\frac{1}{2}\sin \left( 2\alpha \right) {\bar{D}}_{a}\widehat{%
n}^{k}\biggl)dz^{a}, \\
\mathcal{S}_{~\mu }^{\nu }R_{\nu }^{k}dx^{\mu }=& (D\alpha )^{2}\left(
D_{A}\alpha \right) \widehat{n}^{k}dy^{A} +r^{-2}\sin ^{2}\alpha \biggl(\sin ^{2}\alpha \delta ^{sk}\varepsilon
_{ijs}\widehat{n}^{i}{\bar{D}}_{a}\widehat{n}^{j}+\frac{1}{2}\sin \left(
2\alpha \right) {\bar{D}}_{a}\widehat{n}^{k}\biggl)dz^{a},
\end{align}%
\end{widetext}
we obtain 
\begin{align}
\nabla ^{\mu }\left[ R^{\nu },F_{\mu \nu }\right] ^{k}=&4r^{-2}\biggl[%
2(D^{2}\alpha )\sin ^{2}\alpha\nonumber \\
& +\biggl((D\alpha )^{2}-r^{-2}\sin ^{2}\alpha %
\biggl)\sin \left( 2\alpha \right) \biggl]\widehat{n}^{k}
\end{align}%
and finally the Skyrme equations (\ref{nonlinearsigma1}) reduce to the
following single scalar equation on $M^{2}$: 
\begin{align}
0=&(1+2\lambda r^{-2}\sin ^{2}\alpha )D^{2}\alpha
+2r^{-1}(D^{A}r)(D_{A}\alpha ) \nonumber \\
&-r^{-2}\sin \left( 2\alpha \right) \biggl[1-\lambda \biggl((D\alpha )^{2}-r^{-2}\sin ^{2}\alpha \biggl)\biggl]. \label{master1}
\end{align}%
Equations~(\ref{Tmunu}), (\ref{leftEFE}), and (\ref{master1}) give a
complete set of the basic equations in this system.

\subsection{Exact monopole black hole}

The simplest nontrivial solution of the master equation (\ref{master1}) is $%
\alpha =\pi /2+N\pi $, where $N$ is an integer. The energy-momentum tensor (\ref%
{Tmunu}) then becomes 
\begin{align}
T_{\mu \nu }dx^{\mu }dx^{\nu }=&-Kg_{AB}r^{-2}\biggl(1+\frac{1}{2}\lambda
r^{-2}\biggl)dy^{A}dy^{B} \nonumber \\
&+\frac{1}{2}K\lambda r^{-2}\gamma _{ab}dz^{a}dz^{b}.
\end{align}%
It is shown that the most general solution with $\alpha =\pi /2+N\pi $ and $%
(Dr)^{2}\neq 0$ is given by 
\begin{align}
ds^{2}=& -f(r)dt^{2}+f(r)^{-1}dr^{2}+r^{2}(d\theta ^{2}+\sin ^{2}\theta
d\phi ^{2}),  \label{monopoleBH} \\
f(r):=& 1-8\pi GK-\frac{2GM}{r}+\frac{4\pi GK\lambda }{r^{2}}-\frac{1}{3}%
\Lambda r^{2}.
\end{align}%
This solution with $\lambda =0$ (without the Skyrme term) was obtained in 
Ref.~\cite{gibbons1991} and represents a global monopole inside a black hole. In
the present solution, there is the Skyrme contribution in the metric
function which, at first glance, is similar to the Maxwell term in the
Reissner-Nordstr\"{o}m solution. However, unlike the Maxwell case, the
coefficient of the $1/r^{2}$ term \textit{is not an integration constant}
since it is fixed by the couplings of the theory. (This is similar to the
case of the meron black hole~\cite{MBH}). To the best of the authors'
knowledge, the above solution has not been mentioned in any literature. The
metric (\ref{monopoleBH}) with $M=\Lambda =\lambda =0$ is the same as the
Barriola-Vilenkin monopole spacetime~\cite{bv1989}.

It is noted that there are also nonspherical exact solutions with $\alpha=\pi/2+N\pi$ such as the following Taub-NUT-type solution,
\begin{widetext}
\begin{align}
ds^{2}=& -F(r)(dt-2n\cos\theta d\phi)^2+F(r)^{-1}dr^2+(r^2+n^2)(d\theta ^{2}+\sin ^{2}\theta d\phi ^{2}),  \label{TNUT} \\
F(r):=& \frac{r}{r^2+n^2}\biggl((1-8\pi G K-2\Lambda n^2 )r-2M+\frac{4\pi GK\lambda+\Lambda n^4-n^2(1-8\pi G K)}{r}-\frac13 \Lambda r^3\biggl),
\end{align}%
where $n$ is the NUT parameter~\cite{tnut}, and the (Euclidean) Eguchi-Hanson-type solution,
\begin{align}
ds^{2}=& g(r)\frac{r^2}{4}(dt+\cos\theta d\phi)^2+g(r)^{-1}dr^2+\frac{r^2}{4}(d\theta ^{2}+\sin ^{2}\theta d\phi ^{2}),  \label{EH} \\
g(r):=&1-8\pi GK-\frac{32\pi GK\lambda}{r^2}-\frac{a}{r^4}-\frac16 \Lambda r^2,
\end{align}%
where $a$ is a constant.
The solution (\ref{TNUT}) with $n=0$ coincides with our monopole black-hole solution (\ref{monopoleBH}) and the solution (\ref{EH}) with $K=\Lambda=0$ becomes the Eguchi-Hanson space~\cite{eh}.
\end{widetext}

Now we discuss the properties of the spacetime (\ref{monopoleBH}) with $\Lambda
=0$ for simplicity. Although this solution can represent a black hole, the
spacetime is not asymptotically flat but asymptotic to the global monopole
spacetime for $K\neq 0$. The location of the Killing horizon is given by $%
f(r_{\mathrm{h}})=0$, which is solved to give 
\begin{equation}
r_{\mathrm{h}}=\frac{GM}{1-8\pi GK}\biggl(1\pm \sqrt{1-\frac{4\pi K\lambda
(1-8\pi GK)}{GM^{2}}}\biggl).  \label{rh}
\end{equation}%
The relation between $M$ and $r_{\mathrm{h}}$ is 
\begin{equation}
M=\frac{1}{2G}\biggl((1-8\pi GK)r_{\mathrm{h}}+\frac{4\pi GK\lambda }{r_{%
\mathrm{h}}}\biggl).
\end{equation}%
In addition to $K>0$ and $\lambda \geq 0$, we also assume $0<8\pi GK<1$ in
order to have an outer Killing horizon defined by $df/dr|_{r=r_{\mathrm{h}%
}}>0$, which coincides with the black-hole event horizon. The location of
the outer Killing horizon is given by Eq.~(\ref{rh}) with the upper sign, and it
satisfies 
\begin{equation}
\frac{1}{r_{\mathrm{h}}^{2}}<\frac{1-8\pi GK}{4\pi GK\lambda }.
\end{equation}%
The temperature of the black hole is given by 
\begin{equation}
T=\frac{1}{4\pi }\frac{df}{dr}\biggl|_{r=r_{\mathrm{h}}}=\frac{1}{4\pi }%
\biggl(\frac{1-8\pi GK}{r_{\mathrm{h}}}-\frac{4\pi GK\lambda }{r_{\mathrm{h}%
}^{3}}\biggl),
\end{equation}%
while the Wald entropy is 
\begin{equation}
S=\frac{1}{4G}A_{\mathrm{h}}=\frac{\pi }{G}r_{\mathrm{h}}^{2}.
\end{equation}

There is a subtle problem about the global mass of this monopole black hole.
Since there is no free parameter except for $M$, the first law must have the
form of $\delta E=T\delta S$ for some global mass $E$. The parameter $M$
coincides with the ADM mass and satisfies $\delta M=T\delta S$ if and only
if $K=0$. On the other hand, Nucamendi and Sudarsky showed that if the
spacetime approaches to the metric
\begin{align}
ds^{2}=& -g(r)dt^{2}+g(r)^{-1}dr^{2} \nonumber \\
&+(1-\alpha )r^{2}(d\theta ^{2}+\sin
^{2}\theta d\phi ^{2}), \\
g(r)\simeq & 1-\frac{2G{\tilde{M}}}{r},
\end{align}%
${\tilde{M}}$ is identified as the global mass in the monopole spacetime~%
\cite{ns1997}. For our monopole black-hole spacetime, the Nucamendi-Sudarsky
mass is ${\tilde{M}}=M/(1-8\pi GK)^{3/2}$ and--as can be seen directly--it does not satisfy the first law. Instead, by integrating $\delta E=T\delta
S$, we obtain the following expression of $E$: 
\begin{equation}
E=\frac{1}{2G}\biggl((1-8\pi GK)r_{\mathrm{h}}+\frac{4\pi GK\lambda }{r_{%
\mathrm{h}}}\biggl)+E_{0},  \label{monopenergy}
\end{equation}%
where $E_{0}$ is a constant and $M=E-E_{0}$ is satisfied.

Once the first law is fulfilled, it is possible to discuss the
thermodynamical properties of the present black hole with the above energy.
The heat capacity $C$ and the free energy $F$ read 
\begin{align}
C=& \frac{dE}{dr_{\mathrm{h}}}\biggl/\frac{dT}{dr_{\mathrm{h}}} \nonumber \\
=&\frac{2\pi }{%
G}r_{\mathrm{h}}^{2}\biggl((1-8\pi GK)-\frac{4\pi GK\lambda }{r_{\mathrm{h}%
}^{2}}\biggl) \nonumber \\
&\times \biggl(-(1-8\pi GK)+\frac{12\pi GK\lambda }{r_{\mathrm{h}}^{2}}%
\biggl)^{-1}, \\
F=& E-TS \nonumber \\
=&\frac{1}{4G}\biggl((1-8\pi GK)r_{\mathrm{h}}+\frac{12\pi GK\lambda 
}{r_{\mathrm{h}}}\biggl)+E_{0}.
\end{align}%
Although it is difficult to discuss the global thermodynamical stability due
to the fact that we have no {\it a priori} argument to fix the integration
constant $E_{0}$ in Eq.~(\ref{monopenergy}), the local thermodynamical
stability can be analyzed. It is seen that $C<0$ is satisfied for 
\begin{equation}
r_{\mathrm{h}}^{2}>\frac{12\pi GK\lambda }{1-8\pi GK},
\end{equation}%
while $C>0$ holds for 
\begin{equation}
\frac{4\pi GK\lambda }{1-8\pi GK}<r_{\mathrm{h}}^{2}<\frac{12\pi GK\lambda }{%
1-8\pi GK}.
\end{equation}%
This result shows the local thermodynamical stability of a small monopole
black hole with the Skyrme term. Without the Skyrme term, we have $C<0$ and
the black hole is always thermodynamically unstable.


\section{Generalized hedgehog ansatz}

\label{sec4}

\subsection{The ansatz}

In this section, we propose a generalization of the hedgehog ansatz for
self-gravitating Skyrme fields and derive the basic equations in a covariant
form. We start from the following configuration:
\begin{equation}
Y^{0}=\cos \alpha \ ,\ \ Y^{i}=\widehat{n}^{i}\sin \alpha ,
\label{general-Y}
\end{equation}%
which is the same as the hedgehog ansatz, and then $R_{\mu }^{k}$ is given
by 
\begin{equation}
R_{\mu }^{k}=\sin ^{2}\alpha \varepsilon ^{ijk}\widehat{n}^{i}(\nabla _{\mu }%
\widehat{n}^{j})+\frac{1}{2}\sin (2\alpha )(\nabla _{\mu }\widehat{n}^{k})+%
\widehat{n}^{k}(\nabla _{\mu }\alpha ).
\end{equation}%
We now assume the following form of $\widehat{n}^{i}$: 
\begin{equation}
\widehat{n}^{1}=\cos \Theta \ ,\ \ \ \widehat{n}^{2}=\sin \Theta \ ,\ \ \ 
\widehat{n}^{3}=0,  \label{general-Y2}
\end{equation}%
which satisfy $\delta _{ij}\widehat{n}^{i}\widehat{n}^{j}=1$ and hence Eq.~(%
\ref{standard3}). Here $\alpha $ and $\Theta $ are scalar functions. Using
the above expressions, we obtain $\mathcal{S}_{\mu \nu }$ defined by Eq.~(%
\ref{cuadra1}) as 
\begin{equation}
\mathcal{S}_{\mu \nu }=(\nabla _{\mu }\alpha )(\nabla _{\nu }\alpha )+\sin
^{2}\alpha (\nabla _{\mu }\Theta )(\nabla _{\nu }\Theta )\ 
\end{equation}%
and hence 
\begin{equation}
\mathcal{S}=\sin ^{2}\alpha (\nabla \Theta )^{2}+(\nabla \alpha )^{2}.
\end{equation}%
The energy-momentum tensor for the nonlinear sigma model is then given by 
\begin{align}
T_{\mu \nu }=&K\biggl[(\nabla _{\mu }\alpha )(\nabla _{\nu }\alpha )+\sin
^{2}\alpha (\nabla _{\mu }\Theta )(\nabla _{\nu }\Theta ) \nonumber \\
&-\frac{1}{2}g_{\mu
\nu }\biggl((\nabla \alpha )^{2}+\sin ^{2}\alpha (\nabla \Theta )^{2}\biggl)%
\biggl].  \label{sigmatmunu}
\end{align}

Next let us see the field equations for the nonlinear sigma model. It is
shown that they reduce to a single scalar equation under the following
assumptions: 
\begin{eqnarray}
\nabla ^{2}\widehat{n}^{i} &=&L\widehat{n}^{i}\ ,  \label{sessessea1} \\
(\nabla _{\mu }\Theta )(\nabla ^{\mu }\alpha ) &=&0\ ,  \label{sessessea2}
\end{eqnarray}%
where $L$ is a scalar function. This ansatz for the nonlinear sigma model
was first introduced on flat backgrounds in Ref.~\cite{canfora} with a particular
choice of $\Theta $. From Eq.~(\ref{general-Y2}), the condition (\ref{sessessea1}%
) gives $\nabla ^{2}\Theta =0$ and $L=-(\nabla \Theta )^{2}$ and then we
obtain 
\begin{equation}
\nabla ^{\mu }R_{\mu }^{k}=\biggl((\nabla ^{2}\alpha )-\frac{1}{2}(\nabla
\Theta )^{2}\sin (2\alpha )\biggl)\widehat{n}^{k}.
\end{equation}

In summary, the field equations for the nonlinear sigma model (\ref%
{nonlinearsigma1}) (with $\lambda =0$) have been decomposed into the
following equations for $\Theta $ and $\alpha $: 
\begin{align}
& \nabla ^{2}\Theta =0,  \label{KGTheta} \\
& (\nabla ^{2}\alpha )-\frac{1}{2}(\nabla \Theta )^{2}\sin (2\alpha )=0
\label{sessea4}
\end{align}%
with a constraint, Eq.~(\ref{sessessea2}). The corresponding Einstein equations
are sourced by the energy-momentum tensor (\ref{sigmatmunu}). We call the
set of conditions (\ref{general-Y}), (\ref{general-Y2}), and (\ref%
{sessessea2}) the \textit{generalized hedgehog ansatz for nonlinear sigma
models}. Unlike the standard hedgehog ansatz, it also works in systems
without spherical symmetry, as shown in the following subsections.

At first glance, the simplest nontrivial solution $\alpha =\pi /2+N\pi $
of the field equation (\ref{sessea4}) is very similar to the
Einstein-Klein-Gordon system since the energy-momentum tensor (\ref%
{sigmatmunu}) becomes 
\begin{equation}
T_{\mu \nu }=K\biggl[(\nabla _{\mu }\Theta )(\nabla _{\nu }\Theta )-\frac{1}{%
2}g_{\mu \nu }(\nabla \Theta )^{2}\biggl]
\end{equation}%
and $\Theta $ is governed by Eq.~(\ref{KGTheta}). However, as shown in Sec.~%
\ref{sec:axi}, the present system allows a larger class of solutions than
the Einstein-Klein-Gordon system.

Let us add the Skyrme term to the system under the generalized hedgehog
ansatz. Using the expressions
\begin{align}
\mathcal{S}_{\mu \alpha }\mathcal{S}_{\nu }^{~\alpha }=& \sin ^{4}\alpha
(\nabla \Theta )^{2}(\nabla _{\mu }\Theta )(\nabla _{\nu }\Theta ) \nonumber \\
&+(\nabla
\alpha )^{2}(\nabla _{\mu }\alpha )(\nabla _{\nu }\alpha ), \\
\mathcal{S}^{2}-\mathcal{S}_{\alpha \beta }\mathcal{S}^{\alpha \beta }=&
2\sin ^{2}\alpha (\nabla \Theta )^{2}(\nabla \alpha )^{2},
\end{align}%
we obtain the energy-momentum tensor as 
\begin{align}
T_{\mu \nu }=& K\biggl[(\nabla _{\mu }\alpha )(\nabla _{\nu }\alpha )+\sin
^{2}\alpha (\nabla _{\mu }\Theta )(\nabla _{\nu }\Theta )  +\lambda \sin ^{2}\alpha \notag \\
&\times  \biggl((\nabla \Theta )^{2}(\nabla _{\mu }\alpha
)(\nabla _{\nu }\alpha )+(\nabla \alpha )^{2}(\nabla _{\mu }\Theta )(\nabla
_{\nu }\Theta )\biggl)  \notag \\
& -\frac{1}{2}g_{\mu \nu }\biggl((\nabla \alpha )^{2}+\sin ^{2}\alpha
(\nabla \Theta )^{2} \nonumber \\
&+\lambda \sin ^{2}\alpha (\nabla \Theta )^{2}(\nabla
\alpha )^{2}\biggl)\biggl].  \label{em-skyrme}
\end{align}

Now we derive the Skyrme equations. We will show that they reduce to a
single scalar equation under the assumptions (\ref{sessessea2}) and (\ref%
{KGTheta}) and the following additional conditions: 
\begin{align}
&(\nabla ^{\mu }\nabla ^{\nu }\Theta )(\nabla _{\mu }\Theta )(\nabla
_{\nu}\Theta )=0,  \label{general-Y3-1} \\
&(\nabla ^{\mu }\nabla ^{\nu }\alpha )(\nabla _{\mu }\alpha )(\nabla
_{\nu}\Theta)=0\ .  \label{general-Y3-2}
\end{align}%
From Eqs. (\ref{general-Y}), (\ref{general-Y2}), and (\ref{general-Y3-1}), we
obtain 
\begin{equation}
(\nabla _{\mu }\Theta )(\nabla ^{\nu }\Theta )(\nabla ^{\mu }\nabla _{\nu }%
\widehat{n}^{k})=-(\nabla \Theta )^{4}\widehat{n}^{k}.
\end{equation}%
It is a trivial computation to derive the following expressions: 
\begin{widetext}
\begin{align}
\mathcal{S}R_{\mu }^{k}=& \biggl(\sin ^{2}\alpha (\nabla \Theta
)^{2}+(\nabla \alpha )^{2}\biggl)\biggl(\sin ^{2}\alpha \varepsilon ^{ijk}%
\widehat{n}^{i}(\nabla _{\mu }\widehat{n}^{j})+\frac{1}{2}\sin (2\alpha
)(\nabla _{\mu }\widehat{n}^{k})+\widehat{n}^{k}(\nabla _{\mu }\alpha )%
\biggl), \\
\mathcal{S}_{~\mu }^{\nu }R_{\nu }^{k}=& \sin ^{2}\alpha (\nabla
_{\mu }\Theta )(\nabla ^{\nu }\Theta )\biggl(\sin ^{2}\alpha \varepsilon
^{ijk}\widehat{n}^{i}(\nabla _{\nu }\widehat{n}^{j})+\frac{1}{2}\sin
(2\alpha )(\nabla _{\nu }\widehat{n}^{k})\biggl)+(\nabla _{\mu }\alpha
)(\nabla \alpha )^{2}\widehat{n}^{k},
\end{align}%
from which it follows 
\begin{align}
\nabla ^{\mu }(\mathcal{S}R_{\mu }^{k})=& \biggl[(\nabla _{\mu }\alpha
)\nabla ^{\mu }\biggl(\sin ^{2}\alpha (\nabla \Theta )^{2}+(\nabla \alpha
)^{2}\biggl) +\biggl(\sin ^{2}\alpha (\nabla \Theta )^{2}+(\nabla \alpha )^{2}\biggl)%
\biggl((\nabla ^{2}\alpha )-\frac{1}{2}\sin (2\alpha )(\nabla \Theta )^{2}%
\biggl)\biggl]\widehat{n}^{k}, \\
\nabla ^{\mu }(\mathcal{S}_{~\mu }^{\nu }R_{\nu }^{k})=& \biggl[\nabla ^{\mu
}\biggl((\nabla _{\mu }\alpha )(\nabla \alpha )^{2}\biggl)-\frac{1}{2}\sin
(2\alpha )\sin ^{2}\alpha (\nabla \Theta )^{4}\biggl]\widehat{n}^{k}.
\end{align}%
Finally, the Skyrme field equations (\ref{nonlinearsigma1}) reduce to the
following scalar equation: 
\begin{align}
0=& (\nabla ^{2}\alpha )-\frac{1}{2}\sin (2\alpha )(\nabla \Theta )^{2} 
+\lambda \biggl[(\nabla _{\mu }\alpha )\nabla ^{\mu }\biggl(\sin
^{2}\alpha (\nabla \Theta )^{2}\biggl)+\sin ^{2}\alpha (\nabla \Theta
)^{2}(\nabla ^{2}\alpha )-\frac{1}{2}\sin (2\alpha )(\nabla \alpha
)^{2}(\nabla \Theta )^{2}\biggl].  \label{skyrme-eq}
\end{align}
\end{widetext}

In summary, the set of conditions in Eqs. (\ref{general-Y}), (\ref%
{general-Y2}), (\ref{sessessea2}), (\ref{general-Y3-1}), and (\ref%
{general-Y3-2}) define the \textit{generalized hedgehog ansatz for Skyrme
models}, under which the Skyrme equations are decomposed into Eqs.~(\ref%
{KGTheta}) and (\ref{skyrme-eq}). Again $\alpha =\pi /2+N\pi $ is a solution
of Eq.~(\ref{skyrme-eq}). In this case, the Skyrme term does not appear
directly in the geometry, as seen in Eq. (\ref{em-skyrme}). This system is also
not equivalent to the Einstein-Klein-Gordon system because of the
constraints (\ref{general-Y3-1}) and (\ref{general-Y3-2}). (See also the
discussion in Sec.~\ref{sec:axi}.)

In the following subsections, we will present several spacetimes with
suitable isometries which are compatible with the generalized hedgehog
ansatz and, in particular, with the constraints (\ref{sessessea2}), (\ref%
{general-Y3-1}), and (\ref{general-Y3-2}) for $\alpha $ and $\Theta $.

\subsection{Spherically, plane, hyperbolically, and cylindrically symmetric
spacetimes}

The metric in the most general spacetime with spherical ($k=1$), plane ($k=0$%
), or hyperbolic ($k=-1$) symmetry is given by 
\begin{equation}
ds^{2}=g_{AB}(y)dy^{A}dy^{B}+r(y)^{2}\gamma _{ab}(z)dz^{a}dz^{b}.
\end{equation}%
We assume $\alpha =\alpha (y)$ and $\Theta =\Theta (y,z)$ in Eqs. (\ref%
{general-Y}), (\ref{general-Y2}), (\ref{general-Y3-1}), and (\ref%
{general-Y3-2}). The canonical coordinates on the submanifold $%
(K^{2},\gamma_{ab})$ are 
\begin{equation}
\gamma _{ab}(z)dz^{a}dz^{b}=d\theta ^{2}+h(\theta )^{2}d\phi ^{2},
\end{equation}%
where $h(\theta )=\sin \theta $, $1$, and $\sinh \theta $ for $k=1,0,-1$,
respectively. The most general energy-momentum tensor compatible with this
symmetry is given by 
\begin{equation}
T_{\mu \nu }dx^{\mu }dx^{\nu }=T_{AB}(y)dy^{A}dy^{B}+P(y)\gamma
_{ab}dz^{a}dz^{b},  \label{em-general}
\end{equation}%
where $P$ is a scalar on $M^{2}$. The compatibility of the energy-momentum
tensor (\ref{em-skyrme}) with the above form requires $\Theta =\Theta (y)$
or $\Theta =\Theta (z)$.

In the case of $\Theta =\Theta (z)$, the conditions (\ref{sessessea2}) and (\ref{general-Y3-1}) are
fulfilled while Eq.~(\ref{KGTheta}) becomes 
\begin{equation}
{\bar{D}}^{2}\Theta =0,
\end{equation}%
where ${\bar{D}}^{2}:={\bar{D}}_{a}{\bar{D}}^{a}$. 
It is still not clear if there exist solutions of the above equation which give the energy-momentum tensor in
the form of Eq.~(\ref{em-general}) and fulfill the condition (\ref{general-Y3-2}).

In the case of $\Theta =\Theta (y)$, Eqs.~(\ref{KGTheta}) and (\ref%
{sessessea2}) become 
\begin{equation}
D^{2}\Theta+2r (D^{A}r )(D_{A}\Theta ) =0,\qquad (D^{A}\alpha )(D_{A}\Theta
)=0,  \label{eq1}
\end{equation}%
where $D^{2}:=D_{A}D^{A}$. There are two interesting solutions of the above
equations which give the energy-momentum tensor in the form of (\ref%
{em-general}). One is the static spacetime
\begin{align}
ds^{2}=& -g_{tt}(\rho )dt^{2}+g_{\rho \rho }(\rho )d\rho ^{2}+r(\rho
)^{2}\gamma _{ab}dz^{a}dz^{b}, \\
\alpha =& \alpha (\rho ),\qquad \Theta =\varpi t
\end{align}%
and the other is the cosmological spacetime 
\begin{align}
ds^{2}=& -g_{tt}(t)dt^{2}+g_{\rho \rho }(t)d\rho ^{2}+r(t)^{2}\gamma
_{ab}dz^{a}dz^{b}, \\
\alpha =& \alpha (t),\qquad \Theta =\varpi \rho ,
\end{align}%
where $\varpi$ is a constant. In both cases, the conditions (\ref%
{general-Y3-1}) and (\ref{general-Y3-2}) for Skyrme fields are fulfilled.

Our ansatz works also in nonrotating cylindrically symmetric spacetimes. We
consider the most general nonrotating cylindrically symmetric space-time,
\begin{equation}
ds^{2}=g_{AB}(y)dy^{A}dy^{B}+r(y)^{2}d\theta ^{2}+s(y)^{2}d\phi ^{2},
\end{equation}%
and assume $\alpha =\alpha (y)$ and $\Theta =\Theta (y,\theta,\phi)$ in
Eqs.~(\ref{general-Y}) and (\ref{general-Y2}). The most general
energy-momentum tensor compatible with this symmetry is given by 
\begin{equation}
T_{\mu \nu }dx^{\mu }dx^{\nu }=T_{AB}(y)dy^{A}dy^{B}+P_{1}(y)d\theta
^{2}+P_{2}(y)d\phi ^{2}, \label{em-general-cyl}
\end{equation}%
where $P_{1}$ and $P_{2}$ are scalars on $M^{2}$. The compatibility of the
energy-momentum tensor (\ref{em-skyrme}) with the above form requires $\Theta =\Theta (y)$, $\Theta =\Theta (\theta )$, or $\Theta =\Theta (\phi )$. 
Actually, the configuration $\Theta=m\phi$ or $\Theta=m\theta$ is compatible with the generalized hedgehog ansatz, namely, it satisfies the conditions (\ref{sessessea2}), (\ref{general-Y3-1}), and (\ref{general-Y3-2}) and gives the energy-momentum tensor in the form of Eq.~(\ref{em-general-cyl}).
In the case of $\Theta =\Theta (y)$, the following configurations are compatible with
or without the Skyrme term: 
\begin{align}
ds^{2}=& -g_{tt}(\rho )dt^{2}+g_{\rho \rho }(\rho )d\rho ^{2}+r(\rho
)^{2}d\theta ^{2}+s(\rho )^{2}d\phi ^{2}, \\
\alpha =& \alpha (\rho ),\qquad \Theta =\varpi t
\end{align}%
and 
\begin{align}
ds^{2}=& -g_{tt}(t)dt^{2}+g_{\rho \rho }(t)d\rho ^{2}+r(t)^{2}d\theta
^{2}+s(t)^{2}d\phi ^{2}, \\
\alpha =& \alpha (t),\qquad \Theta =\varpi \rho.
\end{align}

\subsection{Axisymmetric spacetimes and nontrivial realization of symmetries}

\label{sec:axi} It is shown that the conditions (\ref{sessessea2}), (\ref%
{general-Y3-1}), and (\ref{general-Y3-2}) are satisfied for the following
configuration: 
\begin{align}
ds^{2}=& h_{ab}(v)dw^{a}dw^{b}+g_{AB}(v)dv^{A}dv^{B},  \label{axi1} \\
\alpha =& \alpha (v),\qquad \Theta =\varpi t+m\phi ,  \label{axi2}
\end{align}%
where $\varpi$ and $m$ are constants and $v^{A}=r,z$. $w^{a}=t,\phi $ are
Killing coordinates and $h_{ab}$ is the induced metric on the Killing
leaves, which are the $\left\{ r={\rm const},~z={\rm const}\right\} $ surfaces. An
interesting example of the above metric is the well-known Weyl-Papapetrou
metric for stationary and axisymmetric spacetimes, 
\begin{equation}
ds^{2}=-Ae^{\Omega /2}\left( dt+\omega d\phi \right) ^{2}+\frac{e^{2\nu }}{%
\sqrt{A}}\left( dr^{2}+dz^{2}\right) +Ae^{-\Omega /2}d\phi ^{2}\ ,
\label{weylpapapetrou1}
\end{equation}%
where the metric functions depend only on $r$ and $z$ (see~Ref.\cite{ernst}). 
We will follow the notation in Ref.~\cite{marco}.

Here let us focus on the nonlinear sigma model. The nonzero components of
its energy-momentum tensor are given by 
\begin{widetext}
\begin{align}
T^z_{~z }+T^r_{~r }=& -KA^{-1}e^{\Omega/2}\biggl[-\varpi
^{2}e^{-\Omega}+(m-\varpi \omega)^2 \biggl]\sin^{2}\alpha \ ,  \label{em-1}
\\
T^z_{~z }-T^r_{~r }=&K\sqrt{A}e^{-2\nu}\biggl[(\partial
_{z}\alpha)^2-(\partial _{r}\alpha)^2\biggl] \ , \\
T^\phi_{~t } =&K \varpi A^{-1}e^{\Omega/2}(m-\varpi \omega) \sin ^{2}\alpha
\ , \\
T^t_{~\phi } =&-KmA^{-1}e^{\Omega/2} \biggl[\omega(m -\varpi \omega) +\varpi
e^{-\Omega}\biggl] \sin ^{2}\alpha \ , \\
T^t_{~t }-T^\phi_{~\phi } =&-KA^{-1}e^{\Omega/2}\biggl[\varpi
^{2}e^{-\Omega}+(m+\varpi \omega)(m-\varpi \omega)\biggl] \sin ^{2}\alpha \ ,
\\
T^t_{~t }+T^\phi_{~\phi } =&-K\sqrt{A}e^{-2\nu}\biggl[(\partial
_{r}\alpha)^2+(\partial _{z}\alpha)^2\biggl] \ .
\end{align}%
The relevant combinations of the tensor $\mathcal{G}_{~\nu }^{\mu }:=G_{~\nu
}^{\mu }+\Lambda g_{~\nu }^{\mu }$ are 
\begin{align}
\mathcal{G}^z_{~z }+\mathcal{G}^r_{~r } =&\frac{e^{-2\nu }}{\sqrt{A}}\left(
\triangle A+2\Lambda \sqrt{A}e^{2\nu }\right) \ ,  \label{sundaba3} \\
\mathcal{G}^z_{~z }-\mathcal{G}^r_{~r } =&\frac{e^{-2\nu }}{\sqrt{A}}\biggl[%
\partial _{r}^{2}A-\partial _{z}^{2}A+\frac{1}{8}A\biggl((\partial
_{r}\Omega )^{2}-(\partial _{z}\Omega )^{2}\biggl)  \notag \\
&-\frac{1}{2}Ae^{\Omega }\biggl((\partial _{r}\omega )^{2}-(\partial
_{z}\omega )^{2}\biggl)-2\biggl((\partial _{r}A)(\partial _{r}\nu
)-(\partial _{z}A)(\partial _{z}\nu )\biggl)\biggl]\ ,  \label{sundaba3-2} \\
\mathcal{G}_{~t}^{\phi } =&\frac{e^{-2\nu }}{2\sqrt{A}}\overrightarrow{%
\nabla }\cdot \left( Ae^{\Omega }\overrightarrow{\nabla }\omega \right) \ ,
\\
\mathcal{G}_{~\phi }^{t} =&-\frac{e^{-2\nu }}{2\sqrt{A}}\biggl[A\omega
\triangle \Omega +\omega \left( \overrightarrow{\nabla }A\right) \cdot
\left( \overrightarrow{\nabla }\Omega \right) +2A\omega e^{\Omega }\left( 
\overrightarrow{\nabla }\omega \right) ^{2}  \notag \\
&+\left( 1+\omega ^{2}e^{\Omega }\right) \biggl\{\alpha \triangle \omega +%
\overrightarrow{\nabla }\omega \cdot \left( \overrightarrow{\nabla }A+A%
\overrightarrow{\nabla }\Omega \right) \biggl\}\biggl]\ , \\
\mathcal{G}_{~t}^{t}-\mathcal{G}_{~\phi }^{\phi } =&-\frac{e^{-2\nu }}{\sqrt{%
A}}\biggl[\frac{1}{2}\overrightarrow{\nabla }\cdot \left( A\overrightarrow{%
\nabla }\Omega \right) +\omega \overrightarrow{\nabla }\cdot \left(
Ae^{\Omega }\overrightarrow{\nabla }\omega \right) +Ae^{\Omega }\left( 
\overrightarrow{\nabla }\omega \right) ^{2}\biggl]\ ,  \label{sundaba4} \\
\mathcal{G}_{~t}^{t}+\mathcal{G}_{~\phi }^{\phi } =&\frac{e^{-2\nu }}{\sqrt{A%
}}\left[ \frac{1}{2}\triangle A+2A\triangle \nu +\frac{1}{8}A\left( 
\overrightarrow{\nabla }\Omega \right) ^{2}-\frac{1}{2}Ae^{\Omega }\left( 
\overrightarrow{\nabla }\omega \right) ^{2}+2\Lambda \sqrt{A}e^{2\nu }\right]
\ ,  \label{sundaba5}
\end{align}%
\end{widetext}
where $\triangle =\partial _{r}^{2}+\partial _{z}^{2}$ and $\overrightarrow{%
\nabla }=\left( \partial _{z},\partial _{r}\right) $. Equations~(\ref{em-1}%
)--(\ref{sundaba5}) provide a complete set of the Einstein equations. It is
seen that, in the static case ($\omega=0$), we have $\mathcal{G}_{~\phi}^{t
}=\mathcal{G}_{~t}^{\phi}=0$ and hence the Einstein equations require $%
\varpi m=0$.

The master equation (\ref{sessea4}) for $\alpha=\alpha(r,z)$ is written as 
\begin{align}
&\frac{e^{-2\nu}}{\sqrt{A}}\biggl(A \triangle\alpha+\overrightarrow{\nabla }
A\cdot \overrightarrow{\nabla }\alpha \biggl) \nonumber \\
& +\frac{e^{\Omega/2}}{2A}\biggl(%
\varpi^2e^{-\Omega}-(\varpi \omega-m)^2\biggl) \sin 2\alpha=0\ .
\end{align}
$\alpha=\pi/2+N\pi$ is again a special solution and gives the following
energy-momentum tensor: 
\begin{align}
T^z_{~z }+T^r_{~r }=& -KA^{-1}e^{\Omega/2}\biggl(-\varpi
^{2}e^{-\Omega}+(m-\varpi \omega)^2 \biggl) \ ,  \label{axi-alpha1} \\
T^z_{~z }-T^r_{~r }=&T^t_{~t }+T^\phi_{~\phi } =0 \ , \\
T^\phi_{~t } =&K \varpi A^{-1}e^{\Omega/2}(m-\varpi \omega) \ , \\
T^t_{~\phi } =&-KmA^{-1}e^{\Omega/2} \biggl(\omega(m -\varpi \omega) +\varpi
e^{-\Omega}\biggl) \ , \\
T^t_{~t }-T^\phi_{~\phi } =&-KA^{-1}e^{\Omega/2}\biggl(\varpi
^{2}e^{-\Omega}+(m+\varpi\omega)(m-\varpi \omega)\biggl) \ .
\label{axi-alpha2}
\end{align}

At first glance, the above form of the energy-momentum tensor can
also be realized by a massless Klein-Gordon field. A linear configuration 
\begin{equation}
\psi (t,\phi )=p_{1}t+p_{2}\phi  \label{linearkg}
\end{equation}
certainly solves the Klein-Gordon equation $\kern1pt%
\vbox{\hrule height 0.9pt\hbox{\vrule width
0.9pt\hskip 2.5pt\vbox{\vskip 5.5pt}\hskip 3pt\vrule width 0.3pt}\hrule height
0.3pt}\kern1pt\psi =0$ in the axisymmetric spacetime (\ref{weylpapapetrou1}%
), where $p_{1}$ and $p_{2}$ are constants. This configuration gives the
following energy-momentum tensor: 
\begin{align}
T_{~z}^{z}+T_{~r}^{r}=& -A^{-1}e^{\Omega /2}\biggl(-p_{1}^{2}e^{-\Omega
}+(p_{2}-p_{1}\omega )^{2}\biggl)\ , \\
T_{~z}^{z}-T_{~r}^{r}=& T_{~t}^{t}+T_{~\phi }^{\phi }=0\ , \\
T_{~t}^{\phi }=& p_{1}A^{-1}e^{\Omega /2}(p_{2}-p_{1}\omega )\ , \\
T_{~\phi }^{t}=& -p_{2}A^{-1}e^{\Omega /2}\biggl(\omega (p_{2}-p_{1}\omega
)+p_{1}e^{-\Omega }\biggl)\ , \\
T_{~t}^{t}-T_{~\phi }^{\phi }=& -A^{-1}e^{\Omega /2}\biggl(%
p_{1}^{2}e^{-\Omega }+(p_{2}+p_{1}\omega )(p_{2}-p_{1}\omega )\biggl)\ ,
\end{align}%
which are indeed the same as Eqs.~(\ref{axi-alpha1})--(\ref{axi-alpha2})
with $p_{1}=\sqrt{K}\varpi $ and $p_{2}=\sqrt{K}m$. However, there is a
crucial difference between the present system and the Klein-Gordon system:
unlike the generalized hedgehog ansatz constructed here, the configuration (%
\ref{linearkg}) of the Klein-Gordon field \textit{is not physical}. Indeed,
the configuration with $p_{2}\neq 0$ is not compatible with the axisymmetric
spacetime because the periodic boundary condition $\psi (t,\phi )=\psi
(t,\phi +2\pi )$ is not satisfied. Even in the case with $p_{2}=0$, if one
assumes that the scalar field is observable, the configuration $\psi =p_{1}t$
is not quite realistic due to the obvious unboundedness of $\psi $ for $t
\to \pm \infty$. In contrast, one obtains the same energy-momentum tensor in
which the fields $Y^{i}$ are completely smooth and bounded in the case of
the nonlinear sigma model and automatically satisfy the boundary conditions,
as can be seen in Eqs.~(\ref{general-Y}) and (\ref{general-Y2}).

Thus, the configuration (\ref{axi2}) discloses a new sector of research of
stationary and axisymmetric spacetimes. Such spacetimes have been deeply
analyzed until now and the solution-generating techniques have been
established for the self-gravitating nonlinear sigma models~\cite%
{heusler,heusler1996}. By adopting the powerful techniques introduced in Ref.~\cite%
{ernst}, however, one assumes that the nonlinear sigma model does \textit{not%
} depend on the Killing coordinates. (See also the recent paper~\cite%
{anabalon2012} on exact solutions with this assumption.) Indeed, in such a
case, the corresponding energy-momentum tensor is trivially compatible with
the spacetime symmetry. In the configuration (\ref{axi2}), in contrast, the
nonlinear sigma model (both with and without the Skyrme term) depends on the
Killing coordinates in a nontrivial way such that the energy-momentum
tensor is still compatible with the spacetime symmetry.


\section{Summary and perspectives}

\label{sec5} In the present paper, we have reinvestigated the hedgehog
ansatz for spherically symmetric spacetimes and considered its
generalization for nonspherically symmetric spacetimes for self-gravitating
nonlinear sigma models and Skyrme models. Our main results are broadly
classified into two types.

In Sec.~\ref{sec3}, we derived the basic equations under the hedgehog
ansatz for future investigations in a fully covariant form on the
two-dimensional orbit spacetime under the spherical isometries. We then
obtained several new exact solutions with or without spherical symmetry.
The spherically symmetric solution represents a global monopole inside a black hole and we have briefly discussed its thermodynamical properties.
The Skyrme term in the metric function resembles the Maxwell term but its coefficient is fixed by the coupling constants.

In Sec.~\ref{sec4}, we proposed the generalized hedgehog ansatz. Under
this new ansatz, the field equations reduce to coupled partial differential
equations for two scalar fields $\alpha $ and $\Theta $ with several
constraint equations between them. 
We have presented some particular configurations compatible with the
generalized hedgehog ansatz in physically interesting spacetimes, including
stationary and axisymmetric spacetimes. In those configurations, the Skyrme
fields depend on the Killing coordinates but the corresponding
energy-momentum tensor does not depend on the Killing coordinates. As a
result, they allow one to implement the spacetime symmetries in a nontrivial
way.

For this reason, the field configurations constructed here are quite
different from the usual ones and it is still unknown at present what kind
of solutions they allow. For this purpose, to extend the
solution-generating techniques to this new sector is an important subject.
Also, the generalized hedgehog ansatz is useful to construct black-hole or
regular solutions numerically. Those studies will shed new light on the
nature of Skyrmions.

\subsection*{Acknowledgements}
The authors thank Masato Nozawa for suggestions about the solutions (\ref{TNUT}) and (\ref{EH}).
HM also thanks the Cosmophysics group in KEK for hospitality and
support. This work has been funded by the Fondecyt grants 1120352 (FC) and
1100328, 1100755 (HM) and by the Conicyt grant "Southern Theoretical Physics
Laboratory" ACT-91. This work was also partly supported by the JSPS
Grant-in-Aid for Scientific Research (A) (22244030). The Centro de Estudios
Cient\'{\i}ficos (CECs) is funded by the Chilean Government through the
Centers of Excellence Base Financing Program of Conicyt.


\end{document}